\documentclass[useAMS,usenatbib]{mn2e} 
\usepackage{color}
\usepackage{graphicx}
\usepackage{dcolumn}
\usepackage{bm}
\usepackage{natbib}
\begin{document}

\title[MGC: SMBH mass function] {The Millennium Galaxy Catalogue: The
$M_{\rm bh}$--$L_{\rm spheroid}$ derived supermassive black hole mass
function}

\author[Vika et al.]
{Marina Vika,$^{1,2}$ Simon P.~Driver,$^{1,2}$ Alister W.~Graham$^{3}$ and Jochen Liske$^4$\\
$^1$Scottish Universities Physics Alliance (SUPA)\\
$^2$School of Physics \& Astronomy, University of St Andrews, North Haugh, St Andrews, Fife, KY16 9SS\\
$^3$Centre for Astrophysics and Supercomputing, Swinburne University of Technology, Hawthorn, Victoria 3122, Australia \\
$^4$European Southern Observatory, Karl-Schwarzschild-Str.\ 2, 85748 Garching, Germany
}

\date{Received XXXX; Accepted XXXX}
\pubyear{2006} \volume{000}
\pagerange{\pageref{firstpage}--\pageref{lastpage}}

\maketitle
\label{firstpage}

\begin{abstract}
Supermassive black hole mass estimates are derived for 1743 galaxies
from the Millennium Galaxy Catalogue using the recently revised
empirical relation between supermassive black hole mass and the
luminosity of the host spheroid. The MGC spheroid luminosities are
based on $R^{1/n}$-bulge plus exponential-disc decompositions. The
majority of black hole masses reside between $10^6 \, M_{\odot}$ and
an upper limit of $2\times10^9 \, M_{\odot}$. Using previously
determined space density weights, we derive the SMBH mass function
which we fit with a Schechter-like function. Integrating the black
hole mass function over $10^6< M_{\rm bh}/ M_{\odot} < 10^{10}$
gives a supermassive black hole mass density of ($3.8 \pm 0.6) \times
10^5 \, h^{3}_{70} \, M_{\odot}$ Mpc$^{-3}$ for early-type galaxies and
($0.96 \pm 0.2) \times10^5 \, h^{3}_{70} \, M_{\odot}$ Mpc$^{-3}$ for
late-type galaxies. The errors are estimated from Monte Carlo
simulations which include the uncertainties in the $M_{\rm bh}$--$L$
relation, the luminosity of the host spheroid and the intrinsic
scatter of the $M_{\rm bh}$--$L$ relation. Assuming supermassive black
holes form via baryonic accretion we find that ($0.008\pm0.002) \,
h_{70}^{3}$ per cent of the Universe's baryons are currently locked up
in supermassive black holes. This result is consistent with our
previous estimate based on the $M_{\rm bh}$--$n$ (S{\'e}rsic index)
relation. 
\end{abstract}


\begin{figure*}
\centering
\includegraphics[height=12cm,width=12.0cm]{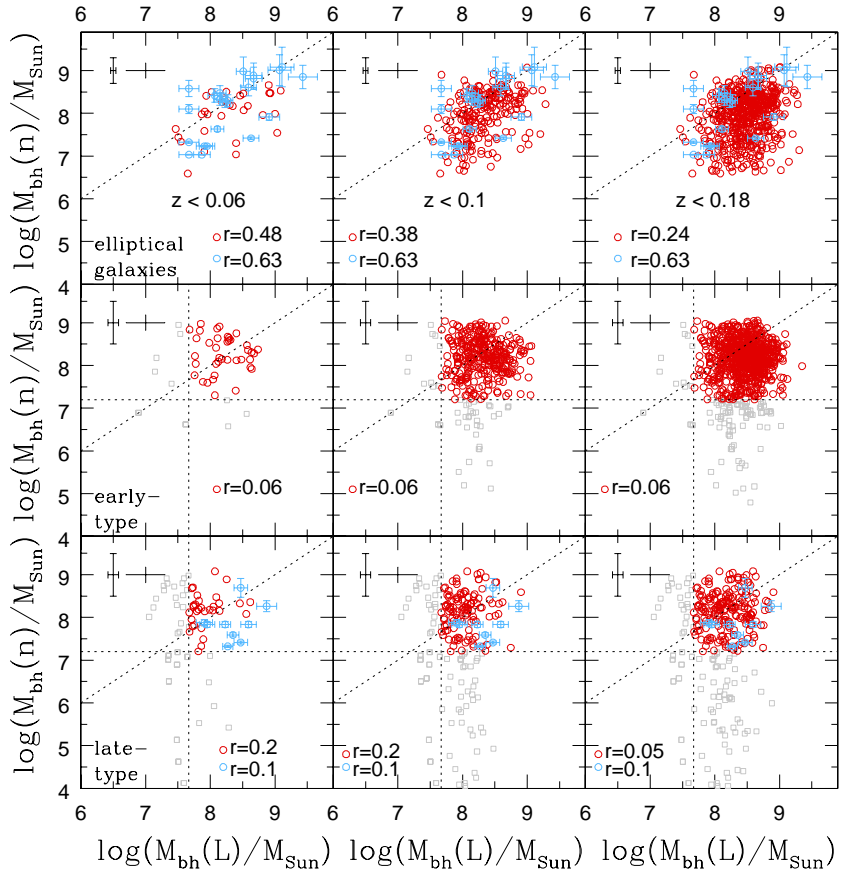}
\caption{A comparison of SMBH masses derived from the $M_{\rm
bh}$--$n$ relation versus the $M_{\rm bh}$--$L$ relation (left panels:
$z < 0.06$, middle panels: $z < 0.1$, right panels: $z < 0.18$). The
red circles are the MGC data. The blue circles denote spheroids from
\citet{tex:CC93}. The spheroids less luminous than $M_{B} = -18$~mag 
or with S{\'e}rsic index $<2$ (indicated with dashed vertical and 
horizontal lines, respectively) are shown as grey
squares as these data are less reliable. The sloped dashed line
indicates the 1-1 relation.  Top panels: (left to right) 30 - 210 - 564
elliptical galaxies (B/T $= 1$), middle panel: 39 - 312 - 867 early-
type disc galaxies ($1 >$ B/T $> 0.4$), bottom panel: 34 - 109 - 312
late-type disc galaxies ($0.4 >$ B/T $> 0.01$). The "right" error bars in each
panel show the average uncertainties on the masses derived from of the 
intrinsic scatters of the $M_{\rm bh}$--$L$,$n$ relations and the "left" 
errors shows those derived from the errors on the spheroid luminosities and S\'ersic index (see Section~3.1). The
correlation coefficient ($r$) has been estimated for each subsample.}
\label{fig:masses}
\end{figure*}

\section{Introduction} 
\label{sec:1}

Observations of nearby galaxies have shown that supermassive black
holes (SMBHs) exist within the centres of both elliptical galaxies and
the classical bulges of disc galaxies (\citealt{tex:KR95,tex:MT98}).
At present there are about 50 credible measurements and 26 partial
measurements as summarised in \citet{tex:G08b}. For lower mass systems
($< 10^8 \, M_{\odot}$), galaxy cores are dominated by a luminous
compact massive object (\citealt{tex:GG03,tex:FC06,tex:BG07}).  
\citet{tex:GT08} found AGN activity in 
early-type galaxies with $M_{\rm gal}< 10^{10} \, M_{\odot}$, revealing that 
SMBHs can exist in low mass galaxies. Furthermore, \citet{tex:SV08} 
found SMBHs can form and evolve in bulgeless galaxies. These argue 
that SMBHs have built up from less massive
`seeds' (e.g., \citealt{tex:WT08}). The SMBH seeds are believed to
form from either PoP III stars collapse (\citealt{tex:VH03}) or directly from dense 
gas collapse (\citealt{tex:KB04}; \citealt{tex:BV06}) in which
large quantities of gas are driven to the galaxy's core giving rise to
an accreting black hole or active galactic nucleus (AGN). Eventually, 
the energy outflow from the AGN may become sufficient to halt further
 gas infall and truncate further star formation in the surrounding stellar
spheroid, either through bright/jet mode or radio/diffuse mode
feedback (e.g.,\citealt{tex:CS06} and \citealt{tex:S08}). 
Although some papers also advocate AGN activity as a trigger for 
star formation \citet{tex:S05} and \citet{tex:PS09}. This
interplay between the SMBH and its surroundings is believed to
give rise to correlations between the SMBH mass and a variety of
measurable properties of the host galaxy's spheroid
component.\footnote{Some properties known to correlate with SMBH mass
are: the spheroid luminosity $L$ (\citealt{tex:KR95,tex:MH03}); the mean
velocity dispersion $\sigma$ of the spheroid
(\citealt{tex:bh4,tex:GB00,tex:G08a,tex:H08}); the spheroid mass
$M_{\rm sph}$ (\citealt{tex:KR95,tex:MH03,tex:HR04}); the galaxy light
concentration $C$ (\citealt{tex:GE01}); the S{\'e}rsic index $n$
of the major-axis surface brightness profile (\citealt{tex:GD07a}). 
See \citet{tex:NF06} for a comparative review.}  After initial formation, 
SMBH growth is believed to progress through periodic gas accretion onto 
the SMBH during major merger events 
(e.g., \citealt{tex:GDz04,tex:DmS05,tex:HH08}).  This
results in the coalescence of the SMBHs and spheroids in such a manner
as to preserve these initial correlations to the present time (\citealt{tex:RC06}). 
Recent evidence does however suggest some
evolution in the SMBH scaling relations and so weakening the criterion
that these relations must be perfectly preserved 
(\citealt{tex:SM06}; \citealt{tex:WT08}; \citealt{tex:SB08}).

AGN activity peaks in the range $1<z<2$ (\citealt{tex:UA03};
\citealt{tex:RS06}; \citealt{tex:HR07}) which sets the key formation epoch for massive
bulge/spheroid formation and the key merger phase of massive dark
matter haloes. If the picture above is correct then the zero-redshift
SMBH mass function and the cosmic AGN energy history should be closely
coupled. In fact, \citet{tex:S81} argued that the AGN cosmic history
and local SMBH mass function can be combined to determine the energy
efficiency of the accretion process. Recent attempts in this direction
have been made by a number of groups (e.g., \citealt{tex:TO06,tex:HH06,tex:HH08};
\citealt{tex:SW07,tex:CL08,tex:YL08}) leading to significant
conclusions. However, the
credibility of these studies rests heavily on the robustness to which
the local SMBH mass function has been measured and to which the cosmic
AGN history is known.

Over recent years there have been a number of estimates of the SMBH
mass function at zero or low redshift and a compendium of some recent
results are presented in Fig.~5 of \citet{tex:SW07}. Except for our
previous work \citep{tex:GD07} none of these studies actually
empirically measure the SMBH mass function but instead estimate it by
analytically combining the measurements of the galaxy luminosity or
velocity function with one of the SMBH scaling relations (as concisely
outlined by \citealt{tex:HR04}, following the first application by
\citealt{tex:SS99}). This inherently incorporates some assumptions of,
for example, the bulge-to-disc ratio and how it varies with luminosity
or how galaxy luminosity relates to galaxy velocity dispersion.

In this paper we depart from this `analytic' approach and pursue a
direct {\it empirical} strategy which uses the recently revised
$M_{\rm bh}$--$L$ relation provided by \citet[][his equations
6~\&~19]{tex:G07} to obtain individual SMBH masses for each galaxy in
the Millennium Galaxy Catalogue (MGC; \citealt{tex:LL03}) which has a
reliable spheroid luminosity. This complements our previous {\it
empirical} method using the $M_{\rm bh}$--$n$ relation
(\citealt{tex:GD07}) and we discuss how the two results compare.

In Section~\ref{sec:2} we describe the data and sample selection. In
Section~\ref{sec:3} we review the latest $M_{\rm bh}$--$L$ relation. In
Section~\ref{sec:4} we use this relation to calculate the SMBH mass
for each galaxy and find the SMBH mass function of local galaxies, the
mass density of SMBHs, the cosmological SMBH density $\Omega_{\rm bh}$
and the baryon fraction of the Universe that is contained in central
SMBHs. Finally, in Section~\ref{sec:5} we compare our results to those
previously derived from the MGC using the $M_{\rm bh}$--$n$ relation
and those recently summarised by \citet{tex:SW07}.

A cosmological model with $\Omega_{\Lambda}=0.7$, $\Omega_{\rm m}=0.3$
and $H_{0}=70$~$h_{70}$~km~s$^{-1}$~Mpc$^{-1}$ is adopted throughout this
paper.

\begin{table*} 
\caption{SMBH mass function data for the full, early- and late-type
samples (DC = dust-corrected), as presented in Fig.~\ref{fig:mf}. The
uncertainties given are the 1$\sigma$ values derived from the Monte
Carlo simulations.}  
\centering
\begin{tabular}{crrrrrr}
\hline
\noalign{\smallskip}
$\log_{10} M_{\rm bh}$ & \multicolumn{6}{c}{$\phi$ $[10^{-4} \, h_{70}^{3} \,
{\rm Mpc}^{-3} \, {\rm dex}^{-1}]$}\\
$[M_{\odot}]$ & All galaxies & Early-type & Late-type &  All galaxies (DC) & Early-type (DC) & Late-type (DC)\\
\noalign{\smallskip}
\hline
\hline
\noalign{\smallskip}
6.25   &    0.00$^{+2.95}_{-0.00}$  &  0.00$^{+0.00}_{-0.00}$ &    0.00$^{+0.00}_{-0.00}$ &  0.00$^{+1.34}_{-0.00}$ &   0.00$^{+0.00}_{-0.00}$ &  0.00$^{+0.00}_{-0.00}$ \\       
6.50   &    1.46$^{+4.79}_{-1.46}$  &  0.00$^{+5.88}_{-0.00}$ &    0.00$^{+2.98}_{-0.00}$ &  0.65$^{+5.44}_{-0.65}$ &   0.00$^{+6.09}_{-0.00}$ &  0.00$^{+1.36}_{-0.00}$ \\
6.75   &    6.63$^{+5.27}_{-5.17}$  &  1.80$^{+4.85}_{-1.80}$ &    3.03$^{+4.40}_{-3.03}$ &  4.62$^{+4.80}_{-3.97}$ &   2.68$^{+4.48}_{-2.68}$ &  1.36$^{+2.20}_{-1.36}$ \\
7.00   &    12.9$^{+5.22}_{-5.47}$  &  4.10$^{+3.88}_{-2.64}$ &    8.33$^{+4.15}_{-4.14}$ &  10.1$^{+5.07}_{-4.80}$ &	5.50$^{+4.17}_{-3.84}$ &  4.33$^{+2.51}_{-2.32}$ \\
7.25   &    18.5$^{+4.45}_{-4.38}$  &  5.98$^{+3.03}_{-2.33}$ &    12.3$^{+3.58}_{-3.31}$ &  17.3$^{+5.10}_{-5.13}$ &	9.73$^{+4.63}_{-4.18}$ &  7.28$^{+2.16}_{-2.11}$ \\
7.50   &    22.0$^{+4.30}_{-3.56}$  &  9.02$^{+2.29}_{-2.16}$ &    12.9$^{+3.48}_{-2.94}$ &  25.8$^{+5.05}_{-4.93}$ &	17.1$^{+4.54}_{-4.42}$ &  8.43$^{+2.05}_{-1.81}$ \\
7.75   &    23.5$^{+3.77}_{-3.06}$  &  12.9$^{+2.14}_{-1.95}$ &    10.5$^{+3.09}_{-2.32}$ &  34.1$^{+4.97}_{-4.53}$ &	25.9$^{+4.49}_{-4.37}$ &  8.00$^{+1.88}_{-1.45}$ \\
8.00   &    22.7$^{+3.46}_{-2.74}$  &  15.6$^{+1.88}_{-1.71}$ &    7.01$^{+2.26}_{-1.69}$ &  38.5$^{+4.84}_{-4.17}$ &	32.2$^{+4.05}_{-3.66}$ &  6.38$^{+1.56}_{-1.27}$ \\
8.25   &    18.5$^{+2.95}_{-2.45}$  &  14.8$^{+1.85}_{-1.62}$ &    3.68$^{+1.36}_{-1.03}$ &  34.0$^{+4.92}_{-4.17}$ &	30.1$^{+3.99}_{-3.57}$ &  3.87$^{+1.17}_{-0.95}$ \\
8.50   &    12.2$^{+2.37}_{-2.07}$  &  10.7$^{+1.84}_{-1.68}$ &    1.51$^{+0.66}_{-0.49}$ &  22.8$^{+4.40}_{-4.05}$ &	21.0$^{+3.96}_{-3.61}$ &  1.78$^{+0.69}_{-0.53}$ \\
8.75   &    6.15$^{+1.81}_{-1.69}$  &  5.71$^{+1.64}_{-1.55}$ &    0.47$^{+0.29}_{-0.21}$ &  11.1$^{+3.51}_{-3.20}$ &	10.5$^{+3.35}_{-3.06}$ &  0.61$^{+0.33}_{-0.26}$ \\
9.00   &    2.15$^{+1.17}_{-0.91}$  &  2.07$^{+1.11}_{-0.87}$ &    0.09$^{+0.11}_{-0.07}$ &  3.63$^{+2.09}_{-1.58}$ &   3.51$^{+2.05}_{-1.53}$ &  0.14$^{+0.14}_{-0.10}$ \\
9.25   &    0.50$^{+0.47}_{-0.29}$  &  0.48$^{+0.48}_{-0.28}$ &    0.00$^{+0.04}_{-0.00}$ &  0.78$^{+0.77}_{-0.47}$ &   0.75$^{+0.78}_{-0.45}$ &  0.00$^{+0.06}_{-0.00}$ \\
9.50   &    0.07$^{+0.13}_{-0.04}$  &  0.07$^{+0.13}_{-0.04}$ &    0.00$^{+0.00}_{-0.00}$ &  0.10$^{+0.19}_{-0.07}$ &   0.10$^{+0.19}_{-0.07}$ &  0.00$^{+0.00}_{-0.00}$ \\
\noalign{\smallskip}
\hline
\end{tabular}
\label{table:points} 
\end{table*}

\section{The Millennium Galaxy Catalogue}                 
\label{sec:2}

The Millennium Galaxy Catalogue (MGC; \citealt{tex:LL03}) is a
medium-deep survey ($\mu_{\mathrm{lim}}=26$~mag~arcsec$^{-2}$)
covering a wide region of sky ($37.5$~deg$^2$) in the $B$-band
($4407$~\AA). The survey extends $75$~deg along the equator (from
$9^{\rm h}$ $58^{\rm m}$ to $14{\rm ^h}$ $47^{\rm m}$). The data
frames were obtained using the 4-CCD mosaic Wide Field Camera on the
2.5-m Isaac Newton Telescope. Each CCD has a pixel scale of
$0.333$~arcsec pixel$^{-1}$. The data were taken with a median seeing
FWHM $=1.3$\arcsec.

Using the {\sc SExtractor} package \citep{tex:gen7} a catalogue was
derived containing over one million objects in the range $16 \le B_{
\rm MGC}< 24$~mag. Initially, two catalogues were defined: MGC-BRIGHT
which includes all galaxies with $B < 20$~mag and MGC-FAINT containing
the rest.  For further details on the MGC imaging data and its
analysis see \citet{tex:LL03}. MGC-BRIGHT is comprised of 10095
resolved galaxies and has 96 per cent complete redshift information
(\citealt{tex:DL05}). This sample was decomposed into bulges and discs
with {\sc GIM2D} using an $R^{1/n}$ S{\'e}rsic profile
for bulges and an exponential profile for discs (\citealt{tex:AD06}).
The robustness of the decomposition process has been verified using
duplicate observations (\citealt{tex:AD06}) and by using extensive
simulations (\citealt{tex:CD09}). All data and
data products used in this paper are freely
available.\footnote{http://www.eso.org/$\sim$jliske/mgc/}

\subsection{Sample selection}

For this study we use the online catalogue mgc-gim2d and extract the
following parameters: spheroid absolute magnitude, bulge-to-total
(B/T) flux ratio, half-light radius of the bulge, redshift and weight
(see Section 4).  We convert the absolute magnitudes from $H_0 =
100$~km s$^{-1}$ Mpc$^{-1}$ to $70$~km s$^{-1}$ Mpc$^{-1}$. We
additionally restrict our sample to the redshift range $0.013$ to
$0.18$. As this sample may contain galaxies with nuclear components
(e.g.\ star clusters) that bias the $R^{1/n}$ model, we remove all
`spheroids' with half-light radii and B/T less than $0.333$ arcsec (1
pixel) and $0.01$, respectively (\citealt{tex:GD07}). We note that the
MGC region is overdense by $9.08$ per cent (Hill et al.\ 2009, in
prep.) and we normalise our weights accordingly.

The sample we construct here is identical to sample 3 of
\citet{tex:GD07} which incorporates a colour cut of $(u-r)_{\rm core}>
2.0$. This colour cut isolates the red spheroid population and is
intended to remove the predominantly blue pseudo-bulge systems.
Pseudo-bulges are believed to form by a process distinct to that by
which classical bulges form, and it is not yet clear whether they
contain a SMBH in their centre and, if so, whether they adhere to the
previously cited spheroid-SMBH relations (see \citealt{tex:H08}). The
colour cut adopted here follows the findings in \citet{tex:DA07} of a
clear colour bimodality within spheroidal systems (as also identified by
\citealt{tex:DF07}).  After we apply the above colour cut we divide
the remaining galaxies into two subsamples. The first contains 1431
`early-type' galaxies (B/T $> 0.4$) and the second contains 312 `late-
type' galaxies ($0.01<$ B/T $<0.4$).
 
Finally, we note that \citet{tex:AD06} consider bulges with $M_{\rm B}
> -17$~mag to be less reliable while \citet{tex:GD07} remove all
galaxies fainter than $M_{\rm B} = -18$~mag from their sample. In this
paper we indicate the `unsafe' area with a dashed vertical line at
$\log(M_{\rm bh}/\rm M_{\odot}) = 7.67$ in all relevant figures which
corresponds to $M_{\rm B} = -18$~mag. Data fainter than this limit
should be treated with caution.

\section{The $M_{\rm bh}$--$L$ relation}
\label{sec:3}

The first empirical correlation between the mass of the SMBH and the
bulge luminosity was identified in the review paper by
\citet{tex:KR95}. Following this initial study a lot of similar
relations have been derived but with differing slopes for the $M_{\rm
bh}$--$L$ relation. As a result, for the same bulge luminosity we have a
range of possible SMBH masses. This large degree of scatter that has
appeared in the luminosity relation is exacerbated by the
difficulties in estimating the bulge luminosity in mid-type
galaxies.

\citet{tex:G07} reviewed the relations of four studies and updated 
their samples with new data (including revised estimates of the distances, 
Hubble type classifications and SMBH masses), added new galaxies, and 
removed some which still have significant uncertainty in their listed 
values. In this paper we use the best fit from a 22 galaxy $B$-band sample 
(\citealt[][his equation 19]{tex:G07}):
\begin{equation}
\log(M_{\rm bh}/M_{\odot})= -0.40(\pm 0.05) (M_{B}+19.5) +
8.27(\pm0.08),
\label{eq:Graham}
\end{equation}
with a total scatter of $0.34$~dex in $\log M_{\rm bh}$ and an
intrinsic scatter of $0.30$~dex. These 22 galaxies were derived by
\citet{tex:G07} from a parent sample of 27 galaxies from
\citet{tex:MH03} Group I galaxies. It is important to note that this
relation does not contain any corrections for internal dust
attenuation. However, recently we demonstrated that dust attenuation
can amount to as much as a few magnitudes in $B$ depending on
inclination (\citealt{tex:DP08}). This poses somewhat of a dilemma as we
have a choice of whether to use the final relationship presented by
Graham and our original uncorrected MGC magnitudes and space densities
(as used in our previous estimate of the SMBH mass function) or
whether to use the dust-free $M_{\rm bh}$--$L$ relation for elliptical
galaxies (\citealt[][his equation~6]{tex:G07}) combined with our dust
corrected MGC bulge magnitudes and the correspondingly revised space
densities. We choose to use and show the results of both alternatives
in order to indicate the possible uncertainty introduced by dust
attenuation.

\begin{figure}
\centering
\includegraphics[height=12cm,width=9.0cm]{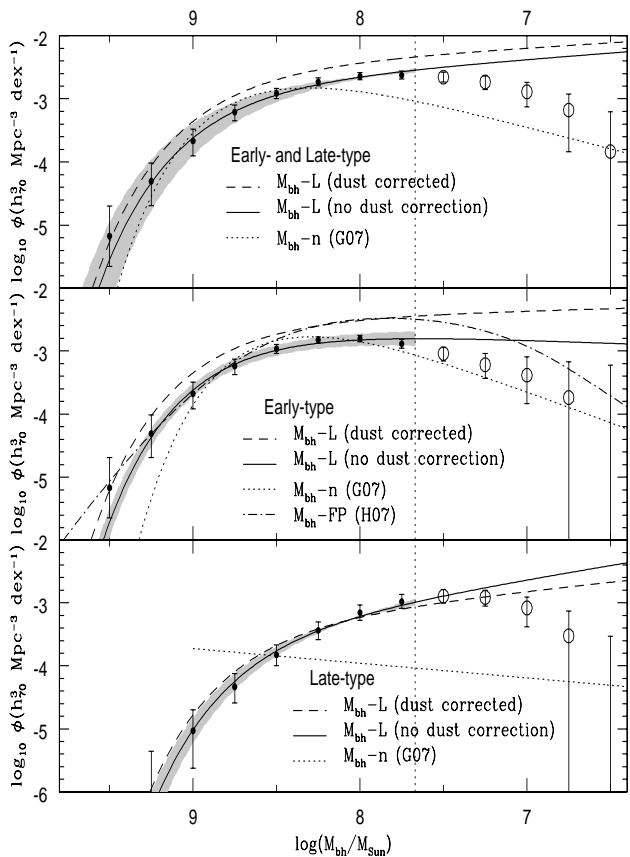} 
\caption{The data points in the top panel show the SMBH mass function
of the full MGC sample. The error bars were derived from Monte Carlo
simulations (see text for details). The solid line and grey shaded
region represent the best fit Schechter function and its
uncertainty. The dashed line shows the same for dust corrected
magnitudes and weights, while the dotted line shows the result of
\citet{tex:GD07} using the $M_{\rm bh}$--$n$ relation (corrected for
the $9.08$ per cent overdensity of the MGC). The middle and bottom
panels show the same for our early- (B/T $> 0.4$) and late-type
($0.01<$ B/T $< 0.4$) samples.  The dashed vertical line at
$\log(M_{\rm bh}/ M_{\odot}) = 7.67$ (corresponding to $M_B =
-18$~mag) indicates our reliability limit and we show all data beyond
this limit with big open circles. The fits use only the reliable data
above this limit. All data points are listed in Table \ref{table:points}. 
The additional curve in the middle panels is from \citet{tex:HH07} (H07).}
\label{fig:mf}
\end{figure}

\begin{figure}
\centering
\includegraphics[height=8cm]{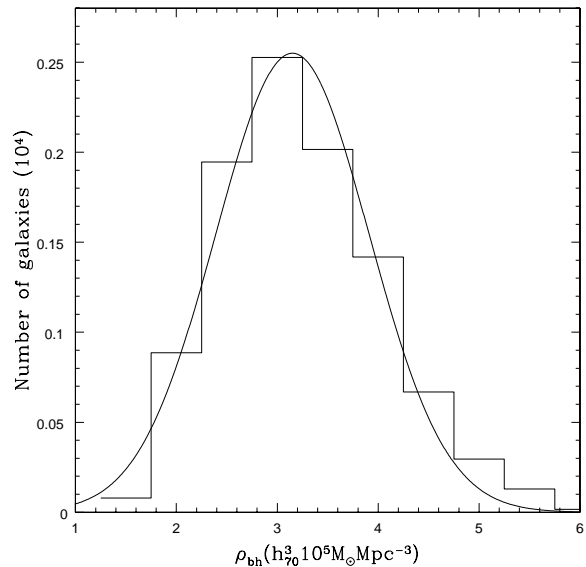} 
\caption{A histogram of the 10001 values of mass density for early-type 
(B/T $> 0.4$) galaxies from the Monte Carlo simulations. The
distribution is reasonably reproduced by a Gaussian (solid line).}
\label{fig:md}
\end{figure}

\begin{table*} 
\caption{Results of a three-parameter Schechter function fit to the
SMBH mass function, both for dust corrected (DC) and uncorrected
samples. The uncertainties on the parameters were derived from Monte
Carlo simulations (see text for details).}
\smallskip
\centering
\begin{tabular}{lcccc}
\hline
\noalign{\smallskip}
Sample & No.\ of & $\log \phi_{\ast}$ & $\log(M_{\ast}/M_{\odot}$) & $\alpha$\\
& spheroids & $[h^{3}_{70}$ Mpc$^{-3} \rm dex^{-1}]$ & &  \\
\noalign{\smallskip}
\hline
\hline
\noalign{\smallskip}
All galaxies (B/T $> 0.01$) (DC) & 1743 & $-3.15~\pm~0.05$   & $8.71~\pm~0.08$  & $-1.20~\pm~0.10$    \\
All galaxies (B/T $ > 0.01$)             & 1743 & $-2.92~\pm~0.05$   & $8.71~\pm~0.06$  & $-1.07~\pm~0.05$    \\
Early-type (B/T $> 0.4$) (DC)  	     & 1431 & $-3.10~\pm~0.05$   & $8.64~\pm~0.03$  & $-0.90~\pm~0.06$    \\
Early-type (B/T $> 0.4$)  	                     & 1431 & $-2.90~\pm~0.04$   & $8.71~\pm~0.03$  & $-1.16~\pm~0.10$    \\
Late-type ($0.01 <$ B/T $ < 0.4$) (DC)     & 312  & $-3.70~\pm~0.02$   & $8.46~\pm~0.03$  & $-1.44~\pm~0.11$    \\
Late-type ($0.01 <$ B/T $< 0.4$)	             & 312  & $-3.66~\pm~0.02$   & $8.49~\pm~0.08$  & $-1.28~\pm~0.01$    \\
\noalign{\smallskip}
\hline
\end{tabular}                                             
\label{table:fit}
\end{table*}

\subsection{Masses from $M_{\rm bh}$--$L$ versus masses from $M_{\rm bh}$--$n$}

\citet{tex:GD07} derived the SMBH mass of each galaxy in the MGC
sample using the photometric quantity S{\'e}rsic index
\footnote{For more information on the S\'ersic index including a
  comprehensive review see \citet{tex:GD05}.} ($n$). In this paper we
derive the SMBH masses using the spheroid luminosity. In
Fig.~\ref{fig:masses} we directly compare the SMBH masses derived for
each galaxy using the two independent methods ($M_{\rm bh}$--$L$ and
$M_{\rm bh}$--$n$) for various subsamples defined in B/T and redshift
as indicated. A correlation between the two BH mass estimates is
not seen and therefore requires explaining. For each subsample we
measure the linear correlation coefficient $r$ and only the
low-$z$ elliptical systems show evidence for a correlation. For the
bulges of disc galaxies, no correlation is detected.

However, we find that the lack of correlation can be well explained
when one considers the errors from both the measurements (on
$L$ and $n$), the intrinsic scatters of the $M_{\rm bh}$--$L$,$n$
relations, and the limited mass range probed here (see error
indicators on Fig.\ref{fig:masses}). In more detail, equation
(\ref{eq:Graham}) above implies that the typical bulge luminosity
error of $\pm0.1$~mag (see fig.~15 of \citealt{tex:AD06}) corresponds
to a mass error of $\pm0.04$~dex. The internal scatter of the $M_{\rm
  bh}$--$L$, however, is $\pm0.30$~dex.
The S{\'e}rsic index value has a typical measurement error of
$\pm20$\% for the elliptical sample and $\pm35$\% for the early \&
late- type sample. While the internal scatter of the $M_{\rm bh}$--$n$
is $\pm 0.18$~dex (\citealt{tex:GD07}). Hence the scatter
in Fig.~\ref{fig:masses} is simply representative of the combination
of two measurement errors and two intrinsic scatters (with the
measurement error dominating for the $M_{\rm bh}-n$ estimates and the
intrinsic scatter dominating for the $M_{\rm bh}-L$ estimate) coupled
with the relatively narrow range of parameter space probed.

To illustrate this conclusion we have used the same $M_{\rm
  bh}$--$L$,$n$ relations as above to calculate the SMBH masses for a
sample of nearby, low-inclination systems within the Virgo and Fornax
clusters (\citealt{tex:CC93}), and for which the measurement errors
are, by comparison, negligible. Over the full range of parameter space
explored, in this study, these data are known to give relatively tight
correlations. However when overplotted (blue points on
Fig.~\ref{fig:masses}) for the narrow range of parameter space we
probe here, they give similarly low correlation coefficients as the
MGC data. This strongly suggests that the lack of correlation is
mainly due to the limited range of parameter space covered by
the MGC data. The lack of correlation is consistent with the adopted
errors (and which are propagated throughout the analysis). The caveat is
that the analysis remains susceptible to any unknown systematic errors
(e.g., bar-bulge contamination) however the overlap between the MGC
data and local data suggests these systematic errors cannot be a
dominant factor. Finally we note that the dominant error in using the
$M_{\rm bh}-L$ relation comes from the inherent intrinsic scatter in
the $M_{\rm bh}-L$ relation rather than the MGC measurement
error. Conversely the reverse is true in our previous study based on
the $M_{\rm bh}-n$ relation.

\section{The MGC SMBH mass function via $M_{\rm bh}-L$}
\label{sec:4}

We have derived individual black hole masses for each spheroid using
equation (\ref{eq:Graham}). These estimates have three sources of
error. The first is the uncertainty on the parameters defining
equation~(\ref{eq:Graham}). This is a systematic error. The second
source of error is the uncertainty in our spheroid magnitude
measurements. Finally, as noted in the previous section, the $M_{\rm
bh}$--$L$ relation has significant internal scatter. We consider the
latter two as random errors. (Note that the alignment of the
calibration sample with the MGC sample on Fig.~\ref{fig:masses}
implies that there is no obvious systematic error in our flux
measurements.) To model these errors appropriately we run a series of
10001 Monte Carlo simulations. For each simulation we randomly modify
the coefficients of equation~(\ref{eq:Graham}) assuming Gaussian error
distributions.  We then perturb each spheroid flux by a random amount
as is appropriate given the spheroid's individual magnitude error, and
calculate a new set of SMBH masses from the perturbed fluxes using the
perturbed version of equation~(\ref{eq:Graham}). Finally, we randomly
modify each of the SMBH masses according to the internal scatter of
the $M_{\rm bh}$--$L$ relation, again assuming a Gaussian
distribution. Given these 10001 Monte Carlo datasets we estimate the
value and 1$\sigma$ error of any given quantity of interest by
calculating that quantity for each of the datasets and determining the
median and 68 percentile range of the resulting distribution,
respectively. This procedure is similar to that followed by
\citet{tex:GD07}.

We now wish to volume-correct our sample and to compute the SMBH mass
function. To this end we employ the space-density weights derived by
\citet{tex:DP07}: the weight of a spheroid of luminosity $L$ is simply
given by the value of the luminosity function of the appropriate
spheroid type, $\phi(L)$, divided by the observed number of spheroids
in the interval $[L,L+{\rm d}L]$. The value of the SMBH mass function
at mass $M_{\rm bh}$ is then calculated as the sum of the weights of
all spheroids within the interval $[M_{\rm bh}, M_{\rm bh}+{\rm
d}M_{\rm bh}]$:
\begin{equation}
\phi(M_{\rm bh})= \sum_{[M,M+{\rm d}M]} W(L) $, where $W(L)= \phi(L)/N(L).
\label{eq:mf}
\end{equation}

In Fig.~\ref{fig:mf} we show the SMBH mass function and its errors as
determined from the Monte Carlo process described above and using
equation~\ref{eq:mf} for our full, early- and late-types subsamples
(top to bottom). All data are listed in Table \ref{table:points}. In
each panel of Fig.~\ref{fig:mf} the solid line represents the best fit
of a three-parameter ($M_{\ast},\phi_{\ast},\alpha$) Schechter-like
function over the mass range $10^{7.75} < M_{\rm bh} / M_{\odot} <
10^{10}$:
\begin{equation}
\phi(M_{\rm bh}) = \phi_{\ast} \left(\frac{M_{\rm bh}}{M_{\ast}}
\right)^{\alpha+1} \exp\left[1-\left(\frac{M_{\rm bh}}{M_{\ast}} 
\right) \right].
\label{eq:schechter}
\end{equation}
The best fitting parameters are given in Table \ref{table:fit}.

Finally, we integrate our Schechter-like estimation of the SMBH mass
function over the mass range probed by our data to calculate the SMBH
mass density, given by the expression:
\begin{eqnarray}
\label{eq:density}
\lefteqn{\rho_{\rm bh} = \int_{\log(M_{\rm bh}/M_{\odot)} = 6}^{\log(M_{\rm
    bh}/M_{\odot)}=10} \phi(M_{\rm bh}) M_{\rm bh} \, {\rm d} \log M_{\rm bh}}
\nonumber\\
& = & \frac{\phi_{\ast} M_{\ast}e^{1}}{\ln(10)} \left[\gamma \left(a+2,
\frac{10^{10} M_{\odot}}{M_{\ast}} \right)  - \gamma
    \left(a+2,\frac{10^6 M_{\odot}}
{M_{\ast}}\right)\right].                          
\end{eqnarray}
As an example of the Monte Carlo process, Fig.~\ref{fig:md} shows the
distribution of the 10001 SMBH density values. This is reasonably
described by a Gaussian function. The final SMBH density is defined as
the median of this distribution with the $1\sigma$ error given by 68
percentile range. The resulting mass densities of SMBHs with masses
between $10^6 < M_{\rm bh}/M_{\odot} < 10^{10}$ for the full, early-
and late-type samples are given in Table \ref{table:density}. The last
column of Table \ref{table:density} also lists the corresponding
cosmological density parameters $\Omega_{\rm bh} = \rho_{\rm bh} /
\rho_{\rm crit}$.

Integrating the mass function over {\em all} masses we derive slightly
larger values for the densities: $5.06$ ($8.5$), $3.8$ ($6.6$) and
$0.96$ ($0.92$) $\times 10^{5} \, h^{3}_{70} \, M_{\odot} \, {\rm
Mpc}^{-3}$ for the full, early- and late-type samples, respectively,
where the numbers in parentheses refer to the dust-corrected
values. These densities are consistent with the values reported
by \citet{tex:GD07}. 

As discussed in the introduction and also in \citet{tex:SS04} if one
assumes that SMBHs form via the accretion of baryons alone and using
the cosmological SMBH mass density $\Omega_{\rm bh,total}= (3.7 \pm
0.7) \times 10^{-6} \, h_{70}$ we can estimate the SMBH baryon
fraction.  Given that $4.5 \, h_{70}^{-2}$ per cent of the critical
density is in the form of baryons \citep{tex:gen8} we find that
$(0.008 \pm 0.002) \, h_{70}^{3}$ per cent of the Universe's baryons
are currently locked up in SMBHs at the centres of galaxies. For
comparison, the value that \citet{tex:GD07} found was $(0.007 \pm
0.003) \, h_{70}^{3}$ per cent.

\begin{table*} 
\caption{SMBH mass densities integrated over the mass range
$10^6$--$10^{10}$ for the full, early- and late-type samples as derived
from the $M_{\rm bh}$--$L$ (dust corrected [DC] and uncorrected) and
$M_{\rm bh}$--$n$ relations \citep{tex:GD07}. The differences in the
numbers of spheroids are due to different absolute magnitude cuts
employed. $\alpha$ is the error due to the uncertainties on the
parameters defining equation (\ref{eq:Graham}) and $\beta$ is from the
uncertainty on the bulge magnitudes. $\gamma$ is derived from the MGC
global cosmic variance of 6 per cent for the effective $30.8$~deg$^2$
region with $0.013 <z< 0.18$. Finally, $\delta$ is the error due to
the intrinsic scatters of the $M_{\rm bh}$--$L$ ($0.30$~dex) and
$M_{\rm bh}$--$n$ relations ($0.18$~dex). All densities have
been corrected for the overdensity of the MGC region (9.08 per cent).}
\smallskip
\centering
\begin{tabular}{lcr@{ $\pm$ }r@{ $\pm$ }r@{ $\pm$ }r@{ $\pm$ }rcc}
\hline
\noalign{\smallskip}
Method  &No.\ of & \multicolumn{5}{c}{$\rho_{\rm bh} \pm \alpha \pm \beta \pm \gamma \pm 
\delta$} & $\Omega_{\rm bh}$\\
& spheroids & \multicolumn{5}{c}{$[10^{5} \, h^{3}_{70} \, M_{\odot} \, {\rm Mpc}^{-3}]$}
& $[10^{-6} \, h_{70}]$\\
\noalign{\smallskip}
\hline
\hline
\noalign{\smallskip}
\multicolumn{4}{l}{Full sample (B/T $> 0.01$):}\\ 	 
  $M_{\rm bh}$--$L$	     & 1743 & $4.9$  & $0.7 $ & $0.5 $ & $0.3 $ & $0.1$  & $3.7 \pm 0.7$\\
  $M_{\rm bh}$--$L_{\rm DC}$ & 1743 & $8.0$  & $1.3 $ & $0.4 $ & $0.5 $ & $0.3$  & $6.2 \pm 1.1$\\
  $M_{\rm bh}$--$n$	     & 1543 & $4.0$  & $1.5 $ & $0.06$ & $0.2 $ & $0.04$ & $2.9 \pm 1.1$\\
\multicolumn{4}{l}{Early-type (B/T $> 0.4$):}\\
  $M_{\rm bh}$--$L$	     & 1431 & $3.8$  & $0.6 $ & $0.4 $ & $0.2 $ & $0.1$  & $2.5 \pm 0.3$\\
  $M_{\rm bh}$--$L_{\rm DC}$ & 1431 & $6.5$  & $1.2 $ & $0.3 $ & $0.4 $ & $0.3$  & $4.8 \pm 0.6$\\
  $M_{\rm bh}$--$n$	     & 1352 & $3.1$  & $1.04$ & $0.05$ & $0.2 $ & $0.03$ & $2.3 \pm 0.8$\\
\multicolumn{4}{l}{Late-type ($0.01 <$ B/T $< 0.4$):}\\
  $M_{\rm bh}$--$L$          & 312  & $0.96$ & $0.2 $ & $0.05$ & $0.06$ & $0.07$ & $0.7 \pm 0.2$\\
  $M_{\rm bh}$--$L_{\rm DC}$ & 312  & $0.92$ & $0.1 $ & $0.04$ & $0.05$ & $0.05$ & $0.7 \pm 0.1$\\
  $M_{\rm bh}$--$n$          & 191  & $0.86$ & $0.49$ & $0.03$ & $0.05$ & $0.02$ & $0.6 \pm 0.4$\\
\noalign{\smallskip} 
\hline
\end{tabular}
\label{table:density}
\end{table*}

\section{Discussion and Comparisons to earlier works} 
\label{sec:5}

\begin{figure*}
\centering
\vspace{-6.0cm}
\includegraphics[width=\textwidth]{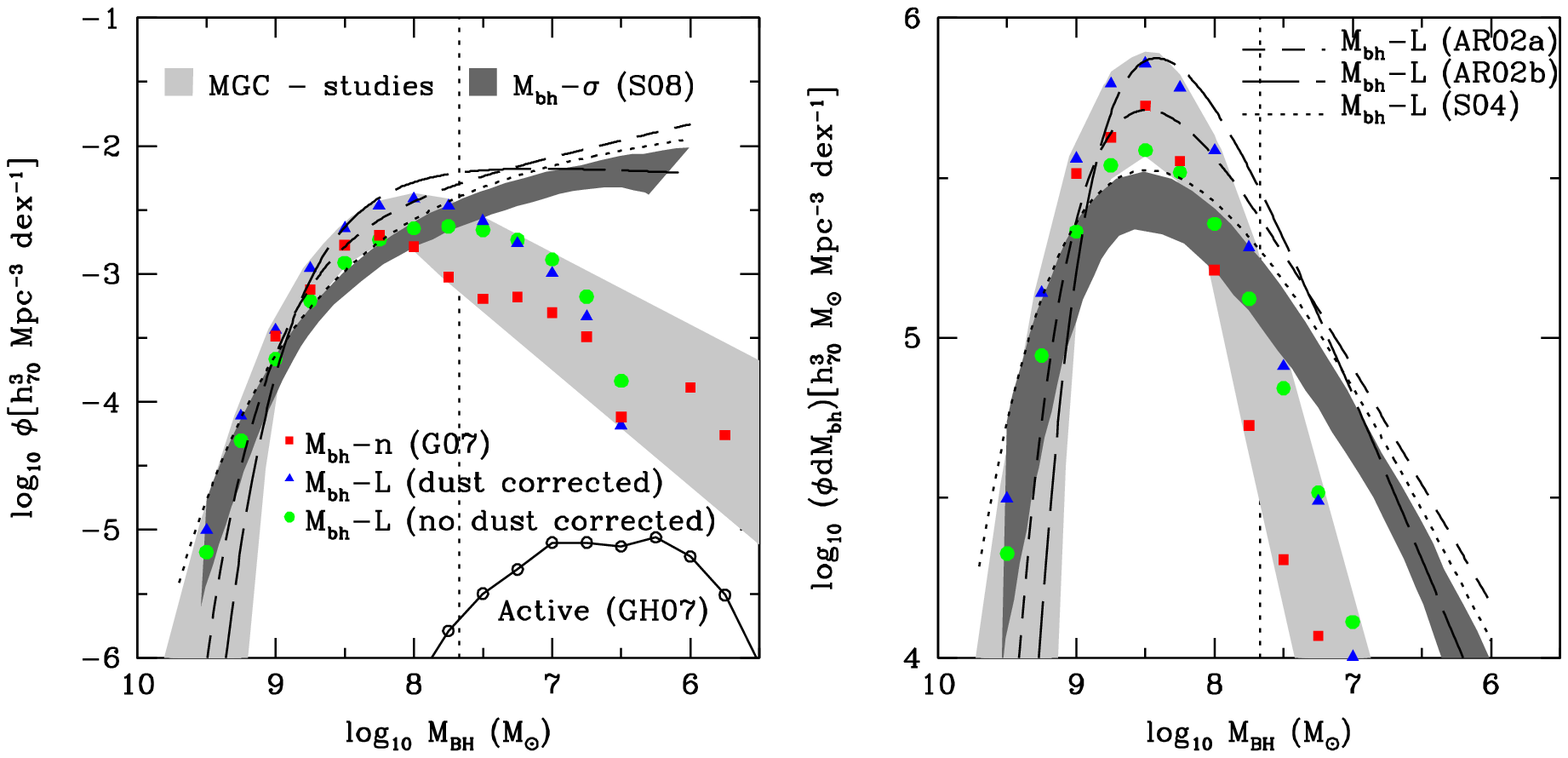} 
\caption{Left panel: a compendium of zero redshift SMBH mass functions
corrected for the hidden Hubble dependency. The curves are from  
\citet{tex:SS04} (S04); \citet{tex:AR02} (AR02a, AR02b);
\citet{tex:GH07} (GH07) and \citet{tex:SB08} (S08,z=0 curve). The light 
grey shaded region represent the MGC results from our three estimates.
The data points show the SMBH mass function for all the MGC subsamples 
(red squares: \citet{tex:GD07}; blue triangles: this study, dust corrected; green 
circles: this study, no dust corrected).
Right panel: equivalent plot showing the contribution by SMBH mass to 
the cosmic SMBH mass density.}
\label{fig:compare2}
\end{figure*}

Fig.~\ref{fig:compare2} shows a compendium of recent estimates of the
zero redshift SMBH mass function in the left panel, while the right
hand panel shows the differential contribution to the mass
density. The shaded (light grey) region representing the MGC results show the
spread in the means from our three estimates as tabulated in
Table~\ref{table:fit} and in Table~1 of \citet{tex:GD07}. Within this
compendium of data only the MGC results are based on actual
measurements of bulge properties. The non-MGC curves/shading all rely
on coupling an empirical SMBH mass-relation with a galaxy luminosity
or velocity function under some assumption as to the fraction of
ellipticals and how the mean bulge-to-total ratio varies with
luminosity. Typically a constant fraction for both values is adopted
independent of luminosity which is contrary to what has been seen in
the MGC and in other morphological studies. In reality the fraction of
ellipticals and the mean B/T ratio should increase with stellar
mass. Incorporating this would have the effect of tilting the mass
functions derived in this `analytical' way, and this is the most
likely explanation for the discrepancy between some of the results
shown in Fig.~\ref{fig:compare2} and those from the MGC (see Appendix A
for details).This echoes
the recent study of nearby active systems by \citet{tex:GH07} who also
find a bounded BH mass function based on an empirical study of the
broad-line regions of 9000 QSOs, albeit shifted to lower black hole
masses (see Fig.~\ref{fig:compare2}, left).

However, it is also fair to note that while the MGC results give a
consistent SMBH mass density and generally agree at the high mass end
they show a broader discrepancy at lower SMBH masses. Most likely this
can be traced back to the bulge-disc decompositions which are no doubt
imperfect, particularly for low-B/T, low-luminosity and/or low-$n$
systems (i.e., those systems likely to contain lower mass
SMBHs). Repeatability tests \citep{tex:AD06} and extensive simulations
(\citealt{tex:CD09}) based on MGC data both show that while disc
parameters are recovered extremely reliably the bulge parameters are
susceptible to gross error. However, no obvious bias via the
simulations has yet been seen, but rather a larger than desirable
general scatter and in this manner the correct answer with appropriate
errors should indeed emerge through the Monte Carlo simulation
process. Note that the modelling of the SMBH mass function includes
extensive Monte Carlo simulations of the structural errors along with
a Malmquist Bias correction.

Within the MGC results the dominant source of random error comes from
the uncertainty in the $M_{bh}$--$n$ and $M_{bh}$--$L$ relations and
the associated intrinsic scatter. This is limited mainly by the sample
size of the relevant calibration datasets. In terms of systematic
error this most likely arises from an, as yet unknown, bias in the
bulge-disc decompositions. While simulations show that systematic
biases are low we do see a discrepancy between the two approaches
which at this stage provides the best and only indication of the scale
of any systematic. Simulations suggest that our bulge luminosity
measurements are more robust than our S\'ersic index measurements
(reflected by the lower final errors in the SMBH mass function) 
and so we therefore consider the new MGC
measurements based on the updated $M_{bh}$--$L$ relation of
\citet{tex:G07}, to be superior to our previous estimate and to the
other `analytical' estimates (given the inherent, sometimes hidden
assumptions in the analytic approach). We therefore advocate the data
presented in this paper as the current best estimate of the zero
redshift SMBH mass function.

SMBH mass functions derived from the M - $\sigma$ relation also show an increase 
in density towards low black hole masses. However velocity dispersions for 
the lower mass systems are generally estimated rather than measured with a 
number of  hidden implicit assumptions required (e.g.,\citealt{tex:AR02}). 
The most recent studies (\citealt{tex:LT07, tex:TM07}) use the 
velocity dispersion function from \citealt{tex:SB03}, however this function 
estimates the black hole masses from the systemic circular velocity which 
could lead to an overestimation of SMBH masses and a steeper faint-end.

\section{Conclusions}
We have used a sample of 1743 galaxies extracted from the Millennium
Galaxy Catalogue to estimate the SMBH masses from the $M_{\rm bh}$--$L$
relation (using \citealt{tex:G07}, his equation 19) to construct the SMBH
mass function at redshift zero. Our results agree well with
our previous study, however, in comparison to other published data the
MGC data consistently show a steep drop-off or decline in the
space-density of SMBHs towards lower masses. Although this lies below
what we consider to be our reliable limits ($ < 10^7 \, M_{\odot}$) it most
likely arises from our more direct empirical approach. Essentially,
the SMBH mass function declines because the spheroid luminosity
function declines (\citealt{tex:DP07}).

Within our reliability limits the contribution of the SMBH mass
function to the total cosmic SMBH density budget is sharply
peaked. Integrating over the best fit mass function we obtain a total
SMBH mass density of $\rho_{\rm bh} = (4.90 \pm 0.7) \times 10^{5} \,
h^{3}_{70} M_{\odot} \, {\rm Mpc}^{-3}$ and a cosmological SMBH
density of $\Omega_{\rm bh} = (3.7 \pm 0.7) \times 10^{-6} \,
h_{70}$. This implies that $(0.008 \pm 0.002) \, h_{70}^{3}$ per cent
of the Universe's baryons are contained in SMBHs, in excellent
agreement with the results from \citet{tex:GD07}.

In this, and our previous, paper we have measured the SMBH mass
function via three methods. Once using the $M_{bh}$--$n$ relation
(\citealt{tex:GD07}) and twice using the $M_{bh}$--$L$ relation (this
paper; once ignoring dust attenuation and once correcting for dust
attenuation). From a comparison of the three SMBH mass functions we
find that the discrepancy in any one bin of the SMBH mass function
(probably the most realistic quantitative insight into the hidden
systematic errors) lies at the 30 per cent level. This constitutes the
current limit which can be achieved from the MGC dataset. A
significant improvement can made in three ways.  Firstly, by
increasing the sample size, secondly by obtaining more robust
measurements, and thirdly by improving the accuracy in which the
$M_{\rm bh}$--$L$,$n$ relations are known. One obvious direction in
the latter case is to re-derive the $M_{\rm bh}$--$L$,$n$ relations in
the NIR to overcome the severe effects of dust attenuation (see
\citealt{tex:DP07}). At the present time no suitable catalogue
exists. Over the coming years we are embarking on a programme to
recalibrate the $M_{\rm bh}$--$L$,$n$ relations at near-IR wavelengths
(where we expect the intrinsic scatter to be significantly reduced),
as well as construct a $\sim100\times$ larger NIR-selected galaxy
sample ($\approx$240k galaxies via the ongoing Galaxy And Mass
Assembly survey; Driver 2008). With these two improvements we should
be able to significantly reduce the errors inherent in this work as
well as probe a significantly broader black hole mass range.

\section*{Acknowledgments}

The Millennium Galaxy Catalogue consists of imaging data from the
Isaac Newton Telescope and spectroscopic data from the
Anglo-Australian Telescope, the ANU 2.3-m, the ESO New Technology
Telescope, The Telescope Nazionale Galileo, and the Gemini Telescope.
The survey has been supported through grants from the Particle Physics
and Astronomy Research Council (UK) and the Australian Research
Council (AUS). We thank the referee for helpful suggestions that
improve this paper. The data and data products are publicly available
from http://www.eso.org/$\sim$jliske/mgc/ or on request from J.~Liske
or S.P.~Driver.

\bibliography{references}

\appendix

\section{Predicting the SMBH Mass function directly from the galaxy luminosity function}
In this paper we have derived the nearby SMBH mass function through
direct estimates of SMBH masses for 1743 individual galaxies selected
from the Millennium Galaxy Catalogue (MGC; see Sample 3 definition
described in \citealt{tex:GD07}). Our resulting SMBH mass function
determined using the $M_{\rm bh}$--$L$ relation is consistent with our
earlier estimate based on the $M_{\rm bh}$--$n$ relation and those
derived analytically as shown in  Fig.~\ref{fig:compare2}. However, below our limit
of bulge reliability (i.e., $M_B > -18$~mag) we note a sharp decline
in our mass function not mirrored in the analytical estimates. In
Section~\ref{sec:5} we claim this is due to inherent assumptions in the
analytical process related to the accuracy to which type fractions are
known and the adoption of a constant bulge-to-total luminosity ratio. To
understand why this is so we firstly describe the analytical method,
we then explore some of the implicit assumptions, and finally derive a
set of analytical SMBH mass functions to illustrate the basis of these
claims. We note that to fully explore this issue it is necessary to
revert to the full MGC sample for which some fraction of our
bulge measurements will be unreliable (see \citealt{tex:CD09} for a
detailed discussion of these issues). We therefore make a strong
caveat that the values and results reported in this appendix are to
qualitatively illustrate some of the subtleties in estimating the SMBH
mass function analytically only and are not sufficiently robust to
produce definitive SMBH mass function estimates at this time.

\subsection{Estimating SMBH mass functions analytically}
Apart from our empirical studies, all estimates of the SMBH mass
function combine a global galaxy luminosity function with estimates of
population ratios and bulge-to-total luminosity ratios. This situation arises
because the Millennium Galaxy Catalogue is the only study, to date, to
actually separate the bulge and disc components for a sufficiently
large sample (\citealt{tex:AD06}). In the analytical approach using,
for example, the $M_{\rm bh}$--$L$ method, a galaxy luminosity function is
adopted (or a set of Hubble type specific luminosity functions) and
the space-density of bulges is then estimated by adopting a constant
bulge-to-total flux ratio (or a set of bulge-to-disc flux ratios for each Hubble
type). These estimated bulge luminosity functions can then be
convolved with the $M_{\rm bh}$--$L$ expression to produce an analytic
SMBH mass function. There are three main issues with this technique:
(i) Hubble population fractions are not well constrained and depend on
luminosity (while type dependent luminosity functions will accommodate
for this, estimates based on the global luminosity function typically
do not). (ii) Bulge-to-total flux ratios are believed to vary with
luminosity (typically all systems fainter than $M_B=-17$~mag are single
component only) and this has not yet been adequately quantified. (iii) A
final more subtle issue is the distinction between the type of bulge,
 simple Hubble classification does not distinguish between
red bulges (classic) and blue bulge (possible pseudo-bulges or
contaminant bars). Our assertion in this paper has been to only
predict SMBH masses for red-bulge systems as it is unclear at this time
whether blue-bulge systems also harbor SMBHs.

\subsection{MGC B/T dependency with luminosity}
Returning to the full publicly available MGC catalogue we can directly
determine the dependency of bulge type and bulge-to-total ratio
as a function of luminosity. First we restrict the sample (to
minimise the contamination by spurious bulges) to the redshift range $0.013<z<0.18$
leaving 7044 galaxies from a parent sample of 10095. We now construct
the following sub-populations (following the arguments laid down in
earlier MGC papers):

\noindent
(1) Red bulge-only (elliptical) systems ($B/T=1$, $(u-r)_c>2$), 579

\noindent
(2) Blue bulge-only (elliptical) systems ($B/T=1$, $(u-r)_c<2$), 235

\noindent
(3) Red bulge systems ($0<B/T<1$, $(u-r)_c>2$), 2487

\noindent
(4) Blue bulge systems ($0<B/T<1$, $(u-r)_c<2$), 964

\noindent
(5) Bulge-less or unresolved bulge systems ($B/T=0$), 2779

\begin{figure}
\centering
\includegraphics[width=\columnwidth]{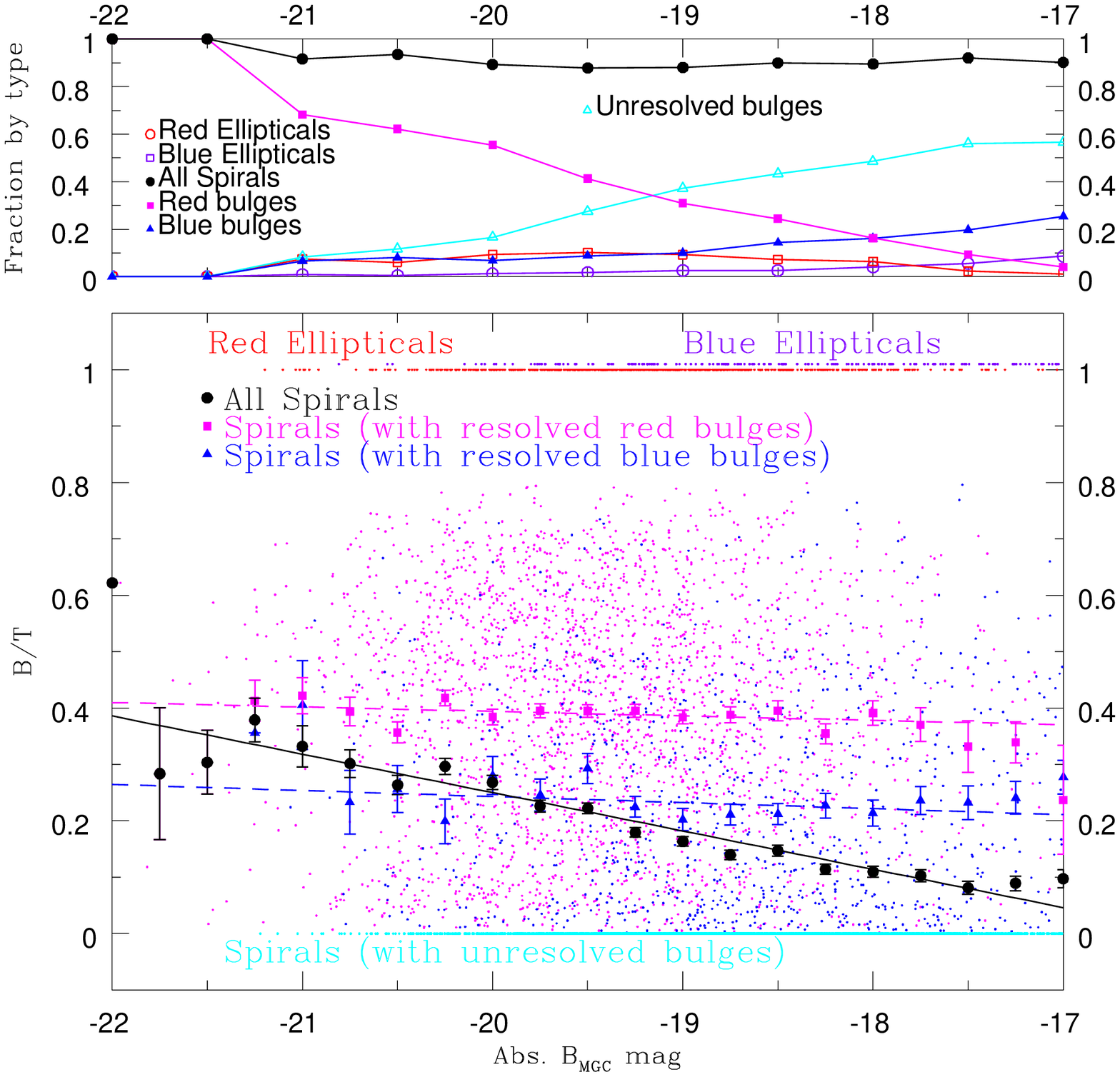}
\caption{({\it main panel}) The distribution of B/T versus absolute
  magnitude for various galaxy sub-populations (as indicated). 
  Overlaid are the mean trends and errors on the mean for
  the red-bulge(solid squares), blue-bulge(solid triangles), 
  and combined spiral population(solid circles).
    ({\it upper}) The relative fraction of various galaxy sub-populations
  colour coded as in the main panel.}
\label{fig:bt}  
\end{figure}

\noindent
We plot these populations on the main panel of Fig.~\ref{fig:bt}. Solid points show
the mean $B/T$ for the red and blue bulge systems as well as that for
the combined spiral sample (red bulge, blue bulge and bulge-less
systems) versus absolute magnitude. We see that the bulge-to-total
fraction of the individual red and blue bulge populations are distinct
but relatively constant. In contrast the overall mean spiral B/T shows
a linear decrease (reasonably well described by the relation
$B/T=-1.09-0.067M_B$).

In the top panel of Fig.~\ref{fig:bt} we show how the fraction of each population varies.
At bright luminosities the sample is dominated by red-bulge systems
which steadily declines, in turn we see a corresponding rise in
blue-bulge and bulge-less systems. The trends seen are quite distinct
showing minimal scatter in the measurement of the means, this suggests
that while errors in some of our bulge measurements are no
doubt present they are unlikely to be dominating.

\subsection{Derivation of the SMBH mass function through scaling relations}
We can now adopt a global galaxy luminosity function (e.g., that
derived in $B$ for the MGC itself in \citealt{tex:DA06}) and combine
this with the scaling relations seen in ~\ref{fig:bt} under various
assumptions as follows:

\noindent
Method 1 --- Adopt a constant $B/T$ and a constant spiral fraction (standard practice).

\noindent
Method 2 --- Adopt a varying $B/T$ and a constant spiral fraction.

\noindent
Method 3 --- Adopt a constant $B/T$ and the red bulge fraction (similar to
our Sample 3).

\noindent
Method 4 --- Adopt a constant $B/T$ and the red and blue bulge fraction.

\begin{figure}
\centering
\includegraphics[width=\columnwidth]{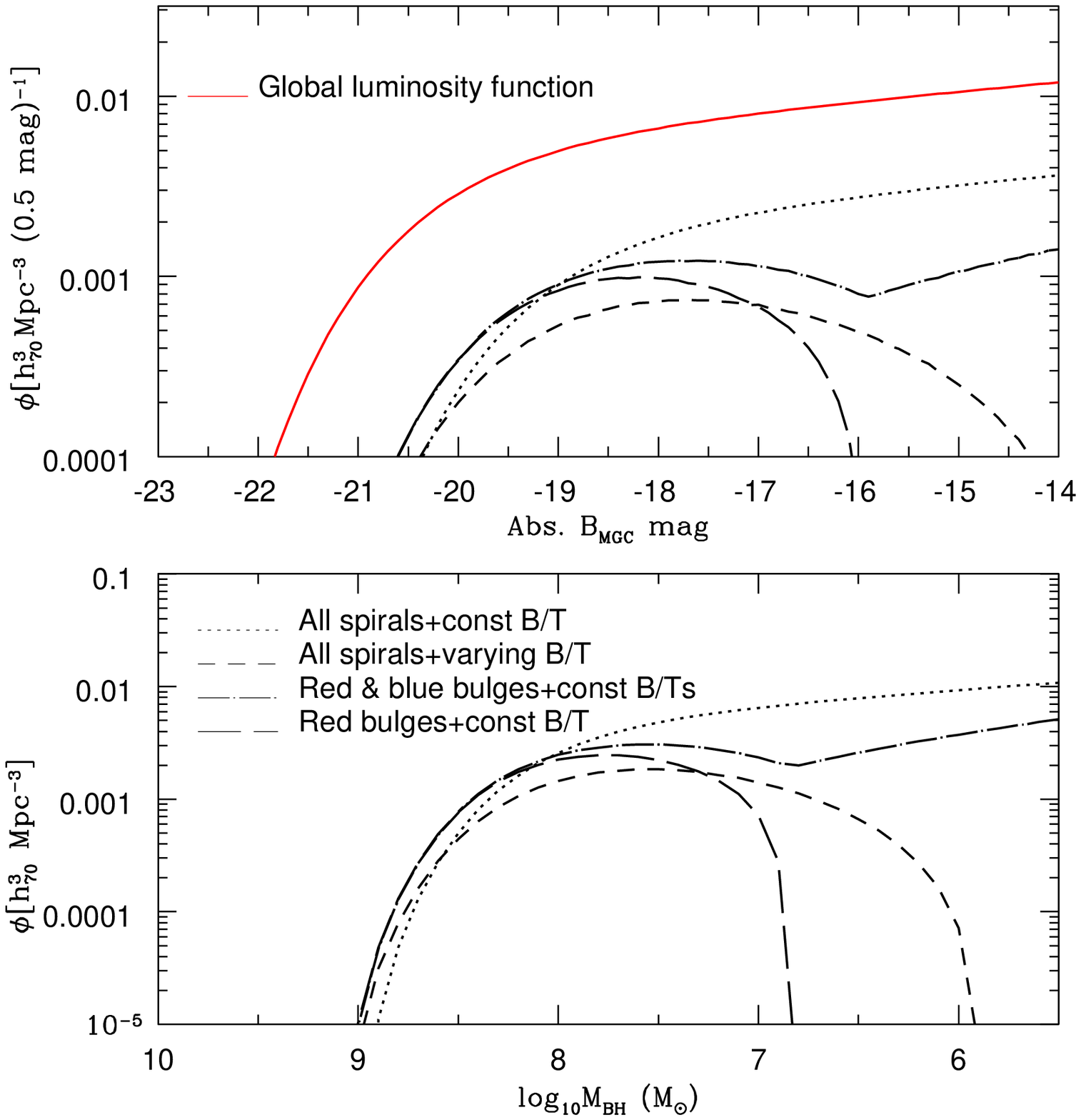}
\caption{({\it upper panel}) The global B-band galaxy
  luminosity function from Driver et al. (2005) and a variety of bugle
  luminosity functions one might derive (as described in the
  text). ({\it lower panel}) The resulting analytical SMBH mass
  functions depending which model LF is adopted.}
\label{fig:lf}
\end{figure}

\noindent
Fig.~\ref{fig:lf} (upper panel) shows the global MGC luminosity function (red/solid) 
and the implied bulge luminosity functions based on these four methods outlined above. 
Fig.~\ref{fig:lf} (lower panel) shows the implied SMBH mass function.
We can see that the faint end of the SMBH mass function
depends critically on these relations and our assumptions. In
particular the key question arises as to whether blue bulges system
(potentially pseudo-bulge systems) harbor SMBHs or not and if so whether they follow 
the same trends as classical (red) bulges. In conclusion,
and having explored both methods, we continue to advocate the
empirical approach laid out in the main body of this paper, where every bulge 
has been measured, and the errors have been derived from Monte Carlo simulations, 
over the analytical approach with its hidden assumptions.

\label{lastpage}

\end{document}